\documentclass[reprint,superscriptaddress,prl,twocolumn,amsmath,amssymb,aps,longbibliography,hidelinks]{revtex4-2}

\usepackage{graphicx}
\usepackage{soul}
\usepackage{dcolumn}
\usepackage{bm}
\usepackage{hyperref}
\usepackage{xr-hyper}
\usepackage{xcolor}
\usepackage[normalem]{ulem}

\usepackage{mathtools}
\newcommand\underrel[3][]{\mathrel{\mathop{#3}\limits_{%
			\ifx c#1\relax\mathclap{#2}\else#2\fi}}}

\begin{document}

\title{Controlling the Glass Transition through Active Fluctuating Interactions}

\author{Emir Sezik}
\email{emir.sezik19@imperial.ac.uk}
\affiliation{Department of Mathematics, Imperial College London, South Kensington, London SW7 2AZ, United Kingdom}

\author{Henry Alston}
\email{henry.alston@phys.ens.fr}
\affiliation{Department of Mathematics, Imperial College London, South Kensington, London SW7 2AZ, United Kingdom}
\affiliation{Laboratoire de Physique, \'Ecole Normale Sup\'erieure, CNRS, PSL Universit\'e, Sorbonne Universit\'e, Universit\'e de Paris, 75005 Paris, France}

\author{Thibault Bertrand}
\email{t.bertrand@imperial.ac.uk}
\affiliation{Department of Mathematics, Imperial College London, South Kensington, London SW7 2AZ, United Kingdom}

\date{\today}

\begin{abstract}
\noindent Fluctuating pairwise interactions are understood to drive fluid-like states in dense biological systems. These states find a broad range of functionalities, such as directing growth during morphogenesis and forming aggregates with heightened mechanical response. However, a tractable model capturing the role of microscopic fluctuating interactions in these structural transitions is crucially lacking. Here, we study a $p$-spin model with fluctuating pairwise couplings (of strength $D_a$ and persistence time $t_a$) as a schematic model for interaction-mediated fluidization. We find that while stronger fluctuations suppress the glass transition, more persistent fluctuations have the opposite effect. We identify the presence of an emergent fluctuation-dissipation relation at long times. We numerically extract the critical temperature $T_c(D_a, t_a)$ from a scaling relation near the transition, illustrating how microscopic fluctuations control the glass transition.
\end{abstract}

\maketitle

The collective fate of many-body systems often hangs on the smallest fluctuations. In the restless world of dense biological matter, interactions never stand still. Instead, pairwise forces are generically subject to fluctuations. For instance, the dynamic turnover of the actin-myosin network controls the adhesion strength between neighbouring cells \cite{Mongera2018, Kim2021}, whereas stochastic type-IV pili mechanics generate intermittent attractive forces between bacteria \cite{Bonazzi2018, Kuan2021}. Across embryonic tissues and cellular aggregates, these fluctuating forces lead to structural transitions between rigid and fluid states which are crucial for controlling tissue growth and remodeling during morphogenesis, and collective motion in health and disease \cite{Sadati2013, Park2015, Garcia2015, Malinverno2017, Bonazzi2018, Mongera2018, Yan2019, Mitchel2020, Henkes2020, Kuan2021, Petridou2021, Grosser2021, Kim2021}; they endow living matter with a remarkable ability to switch between solid-like and liquid-like states going across a glass transition \cite{Angelini2011, Parry2014, Bi2015, Bi2016, Atia2018}. 

Glassy systems are characterized by disordered (liquid-like) yet mechanically rigid (solid-like) structures \cite{Angell1995, Debenedetti2001, Berthier2011, Janssen2018, Mezard1986}. At low temperatures, their relaxation times exceed any laboratory scale, leaving them unable to equilibrate with their environment \cite{Lubchenko2007, Berthier2011, Lubchenko2015}. This arises from frustration among numerous metastable states \cite{Bouchaud1998, Charbonneau2014} and leads to aging and memory effects \cite{Struik1978, Bouchaud1992, Kob1997a, Jonason1998, Bouchaud2000, Cipelletti2000, Dupuis2001, Lunkenheimer2005, Wolynes2009, Scalliet2019, Janzen2024}, dynamical heterogeneity \cite{SchmidtRohr1991, Kob1997b, Weeks2000, Ediger2000, Debenedetti2001, Widmer-Cooper2004, Lubchenko2007, Chamon2007, Berthier2011}, ergodicity breaking \cite{Bouchaud1992, Cugliandolo1995, Jonason1998, Bernaschi2020}, and fluctuation–dissipation violations \cite{Cugliandolo1993, Bouchaud1998, Weeks2000, Crisanti2003, Gotze2008}. To address structural transitions in living systems, however, one must consider active matter. Indeed, at high densities, motile active matter shows phase behavior analogous to, yet distinct from, passive glasses \cite{Berthier2013, Berthier2017, Nandi2018, Janssen2019, Berthier2019, Klongvessa2019, Loewe2020, Mandal2020a, Mandal2020b, Mandal2021, Yang2022, Keta2022, Janzen2022, Wiese2023, Yoshida2024, Anand2024}. Early studies revealed that self-propulsion can shift and reshape the glass transition, either fluidizing dense suspensions or, conversely, promoting kinetic arrest depending on the propulsion mechanism and persistence \cite{Henkes2011, Ni2013, Berthier2013, Berthier2014, Flenner2016}. For non-motile active matter, the framework of kinetic theory has proved successful in describing the emergent structures observed in many-particle systems with active pairwise forces \cite{MonchoJorda2020, Kuan2021, Bley2021, Alston2022a}, but its applicability is ultimately limited at high particle density.

A quantitative theory describing the glass transition remains elusive \cite{Anderson1995}. Even in thermal systems, exact analytical results are largely confined to infinite-dimensional mean-field models, such as the hard sphere glass transition \cite{Charbonneau2014a, Charbonneau2014b, Parisi2010, Parisi2020}. More tractable descriptions arose from mode-coupling theory \cite{Bengtzelius1984, Leutheusser1984, Bouchaud1996, Gotze2008}, derived \textit{from first principles} using the Mori-Zwanzig projection formalism \cite{Szamel1991,Reichman2005}. In active systems, mode-coupling theory successfully describes how self-propulsion alters relaxation pathways and collective dynamics, thus leading to modified glass transition lines \cite{Szamel2016, Nandi2017, Liluashvili2017,  Szamel2019, Debets2022, Debets2023}. Recent works further highlighted that dynamical heterogeneities in active glasses acquire features absent in equilibrium \cite{Paul2023}.

While mode-coupling theory has its own limitations \cite{Berthier2011}, it naturally links structural glass transitions to spin glass theory. The $p$-spin model \cite{Gross1984, Gardner1985}---a generalization of the Sherrington–Kirkpatrick model \cite{Sherrington1975} to higher-order interactions---provided the first mean-field framework capturing the phenomenology of glass formation \cite{Gross1984, Gardner1985}. Its dynamics for $p=3$ map onto mean-field mode-coupling descriptions of supercooled liquids \cite{Kirkpatrick1987a, Kirkpatrick1987b, Kirkpatrick1989} making it a cornerstone of the random first-order transition theory \cite{Biroli2012} and a minimal, analytically tractable setting to study dynamical arrest, metastability, and the Gardner transition in thermal systems \cite{Crisanti1993, Cugliandolo1993, Bouchaud1996, Cugliandolo1997a, Castellani2005, Folena2020, Ghimenti2022}.

Non-equilibrium extensions of the $p$-spin model have since been used to explore these phenomena in driven and active systems \cite{Cugliandolo1997b,  Berthier2000a, Berthier2000b, Berthier2001, Berthier2013, Ghimenti2022}. Adding external driving to each spin, as a proxy for the persistent motion of spins and a minimal source of self-propulsion-like activity, produces either partial or full suppression of the glass transition, depending on the driving timescale relative to glassy dynamics \cite{Berthier2013}. When present, the non-equilibrium transition still shows two-step decay and diverging equilibration time, as in the thermal $p$-spin model. 

In this Letter, we show how active couplings reshape the glass transition in the $p$-spin model. These dynamical couplings are representative of fluctuating microscopic pairwise forces in structural glasses, a bona fide source of activity, requiring a finite rate of energy dissipation \cite{Alston2022b, Cocconi2023b}. This yields an analytically tractable model of a non-equilibrium glass transition controlled by active fluctuating interactions. We analyze both extrinsic and intrinsic interaction fluctuations, characterized by strength $D_a$ and persistence time $t_a$, showing that they can suppress the glass transition and fluidize the system. Furthermore, we demonstrate that they generate two distinct fluctuation--dissipation relations (FDR) valid over different timescales. Altogether, our results reveal a genuinely non-equilibrium glass transition inimitable in thermal $p$-spin models.

\begin{figure*}[t]
    \centering
    \includegraphics[width=\linewidth]{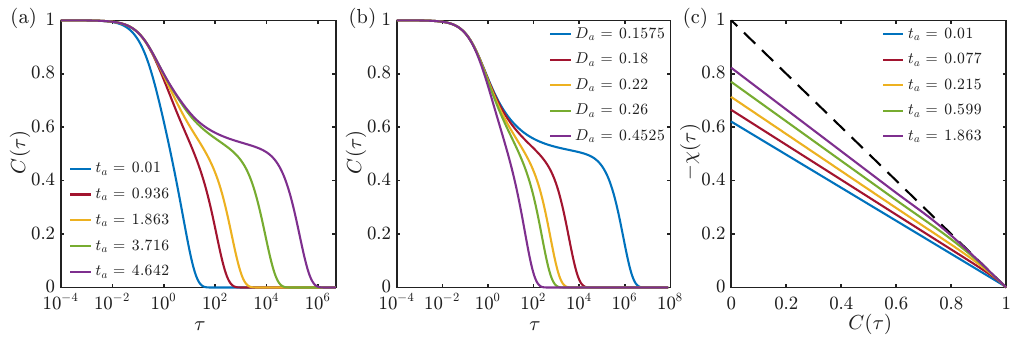}
    \caption{\textit{Correlation functions and effective FDRs in fluid phase ---} (a) Correlation functions $C(\tau)$ for different values of the interaction fluctuation persistence time $t_a$, fixing $D_a = 0.4525$ and $T = 0.535$. (b) Correlation functions $C(\tau)$ shown for different interaction fluctuation strengths, $D_a$ at $T = 0.555 < T_{g}^{\rm eq}$ and $t_a = 0.6$. While increasing the interaction persistence time $t_a$ drives the system towards the glass transition, increasing the fluctuation strength $D_a$ has the opposite effect. (c) Integrated response function $\chi(\tau)$ as a function of the correlation function $C(\tau)$ for $T = 0.535$ and $D_a = 0.4525$. We identify two linear regimes, implying the existence of an effective FDR at both short and long times where the effective temperature is given by the gradient in each regime.}
    \label{fig:shoulder_Da}
\end{figure*}

\textit{Active $p$-spin model with fluctuating couplings.---} We consider an extension of the $p$-spin model supplemented by dynamic $p$-wise couplings in the mean-field limit. The $p$-wise coupling $\eta_{i_1 \cdots i_p}(t)$, unlike the usual quenched coupling $J_{i_1 \cdots i_p}$, fluctuates in time according to dynamics \textit{independent} from the spins. These intrinsic fluctuations break detailed balance and drives the system away from equilibrium. A thermal $p-$spin model is defined by the following Hamiltonian: 
\begin{equation}
    \mathcal{H}[\{\sigma\}] = -\sum_{i_1 < \cdots < i_p}^N J_{ i_1 \cdots i_p} \sigma_{i_1} \cdots \sigma_{i_p} + \mu \sum_{i_1 = 1}^{N} \sigma^2_{i_1}
\end{equation}
where $J_{ i_1 \cdots i_p}$ is a quenched variable drawn from a Gaussian distribution with zero mean and variance $p!/(2N^{p-1})$ and $\mu$ is a Lagrange multiplier enforcing the spherical condition $\sum_{i = 1}^{N} \sigma^2_i = N$. The dynamics of this model under the effect of $p$-wise couplings is governed by the following Langevin equation:
\begin{align}
\label{eq:pspinwithfluctuatingcouplings_Langevin}
    \dot{\sigma}_{i_1}(t) = &-\frac{\partial \mathcal{H}[\{\sigma\}]}{\partial \sigma_{i_1}} \nonumber\\
    	&+ \sum_{i_2< \cdots < i_p} \eta_{i_1\cdots i_p}(t) \sigma_{i_2}(t) \cdots \sigma_{i_p}(t) + \xi_{i_1}(t)
\end{align}
where $\xi_i(t)$ is a Gaussian white noise with zero mean satisfying $\langle \xi_i(t) \xi_j(t') \rangle = 2T \delta_{ij} \delta(t-t')$ and $\eta_{i_1 \cdots i_p}(t)$ is assumed to undergo an OU-like process:
\begin{equation}
\label{eq:pspinwithfluctuatingcouplings_OU}
    t_a \dot{\eta}_{i_1\cdots i_p}(t) = -\eta_{i_1\cdots i_p}(t) + \zeta_{i_1 \cdots i_p}(t)
\end{equation}
with $\zeta_{i_1 \cdots i_p}(t)$ a Gaussian white noise with zero mean satisfying $\langle \zeta_{i_1 \cdots i_p}(t) \zeta_{j_1 \cdots j_p}(t')\rangle = D_a p!/(2N^{p-1}) \delta(t-t') \prod_{k = 1}^p \delta_{i_k, j_k}$. 
Physically, in this model, $D_a$ corresponds to the energy imparted to the system due to active fluctuating couplings and $t_a$, their persistence time, controls the amount of time over which this injection occurs.

Here, we assumed that the $p$-wise fluctuations are independent and local, however, this need not be the case. In \footnote{See Supplemental Material at [] for further analytical and computational details, which includes Refs. \cite{Cugliandolo2003, Barrat1997, Kim2001, Fuchs1991, Flenner2005, Miyazaki2004}}, we also consider (i) the case of extrinsic fluctuations, where the relative strength of fluctuations is the same for all couplings, and show that the resulting effective equations of motion have the same form for both of these models and (ii) the case of a general process only requiring the $\eta_{i_1 \cdots i_p}(t)$ to be drawn from a Gaussian distribution such that $ \langle \eta_{i_1 \cdots i_p}(t) \rangle = 0$ and $\langle \eta_{i_1 \cdots i_p}(t)\eta_{j_1 \cdots j_p}(t')\rangle = D_a p!/(2 N^{p-1})F(t-t')\prod_{k = 1}^p \delta_{i_k, j_k}$, where $F(\tau)$ quantifies the persistence of the noise.

\textit{Dynamical description in the fluid phase. ---} Starting from Eqs. (\ref{eq:pspinwithfluctuatingcouplings_Langevin}) and (\ref{eq:pspinwithfluctuatingcouplings_OU}), one can derive a closed set of equations of motion for the correlation function
\begin{equation}
C(t,t') = \frac{1}{N} \sum_{i = 1}^{N} \langle \sigma_i(t) \sigma_i(t') \rangle
\end{equation}
and the response function
\begin{equation}
R(t,t') = \frac{1}{N} \sum_{i = 1}^{N} \left\langle \frac{\delta \sigma_i(t)}{\delta \xi_i(t')}\right\rangle.
\end{equation}
These two functions fully characterize the dynamics of the system \cite{Note1}. 

For high enough temperature $T$, the system is in a fluid phase where we can assume that the dynamics are stationary for long enough waiting times $t'$, i.e. $C(t,t') = C(t-t')$ and $R(t,t') = R(t-t')$ as $t' \rightarrow \infty$. In this regime, the equations of motion are \cite{Note1}
\begin{subequations}
    \label{eq:stationaryeomCandR}
    \begin{align}
         \frac{ \partial C(\tau)}{\partial \tau} &= -\mu C(\tau) + \int_0^{\tau} dt \tau' \Sigma(\tau - \tau') C(\tau')   \\
         &\quad + \int_{0}^{\infty} d\tau' \left[ \Sigma(\tau + \tau') C(\tau') + D(\tau + \tau') R(\tau') \right] \nonumber\\
         \frac{\partial R(\tau)}{\partial \tau} &= -\mu R(\tau) + \int_{0}^{\tau} d\tau' \Sigma(\tau - \tau') R(\tau') +\delta(\tau) \\
         \mu &= T + \int_{0}^{\infty} d\tau' \left[ \Sigma(\tau') C(\tau') + D(\tau') R(\tau') \right]
    \end{align}
\end{subequations}
where 
\begin{subequations}
\begin{align}
\Sigma(\tau) &= \frac{p(p-1)}{2}C^{p-2}(\tau) R(\tau)(1 + D_a F(\tau)), \\
D(\tau) &= \frac{p}{2}C^{p-1}(\tau)(1 + D_a F(\tau)), \\
F(\tau) &= \frac{1}{2t_a}e^{-|\tau|/t_a}
\end{align}
\end{subequations}
and $\mu$ enforces the spherical condition $C(0) = 1$. It can be easily verified that by setting $D_a = 0$, we recover the equations of motion that describe the fluid phase of the thermal $p$-spin model \cite{Cugliandolo1993, Kirkpatrick1987b, Kirkpatrick1987a, Berthier2011}. In deriving the stationary equations, we have neglected the aging contribution, which dominates in the glass phase. Consequently, Eqs. (\ref{eq:stationaryeomCandR}) cease to hold within the glass phase, only capturing the onset of it.

\begin{figure}[b]
    \centering
    \includegraphics[width=1\linewidth]{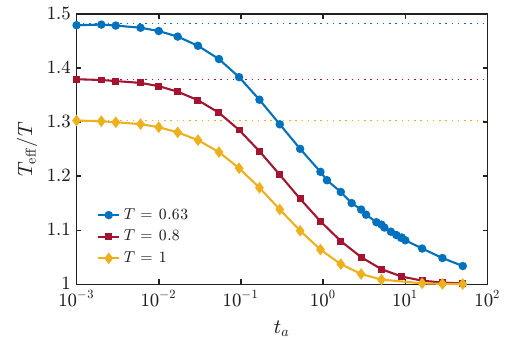}
    \caption{\textit{Measured effective temperature for long time FDR ---} Having identified the long time FDR present in Fig.\,\ref{fig:shoulder_Da}(c), we measure $T_{\rm eff}$ as a function of $t_a$ for $D_a = 0.405$ and varied temperature $T$. The dashed lines denote the analytic result derived for $T_{\rm eff}$ in the white noise limit, i.e. a limit of non persistent fluctuations $t_a\rightarrow0$.}
    \label{fig:T_eff_vs_ta}
\end{figure}

\textit{Suppression of the glass transition by active fluctuations.---} Eqs. (\ref{eq:stationaryeomCandR}) can be fully understood analytically in the zero persistence limit, $t_a \rightarrow 0$, where fluctuations effectively become infinitely fast and $F(\tau) \rightarrow \delta(\tau)$. In this limit, the dynamics are that of a thermal $p$-spin model with effective temperature $T_{\rm eff} = T + pD_a/4$ and the fluctuations purely heat up the system. Furthermore, because of this mapping to a thermal $p$-spin model, we also recover an \textit{effective} FDR, $R(\tau) = -\frac{1}{T_{\rm eff}} \frac{dC(\tau)}{d \tau}$. We use this to write a closed-form equation for $C(\tau)$:
\begin{equation}
\label{eq:whitenoisecase}
    \frac{d C(\tau)}{d \tau} = -T_{\rm eff} C(\tau) - \frac{p}{2 T_{\rm eff}} \int_0^{\tau} d\tau' C^{p-1}(\tau - \tau') \frac{dC(\tau')}{d\tau'}
\end{equation}
which is precisely the equation of motion for a thermal $p$-spin model in the fluid phase. Standard results provide the glass transition temperature, $T_g$, which is present for $p\geq3$ \cite{Cugliandolo1993, Kirkpatrick1987a, Kirkpatrick1987b, Note1}: 
\begin{equation}
     T_g =  \sqrt{\frac{p(p-2)^{p-2}}{2(p-1)^{p-1}}} - \frac{pD_a}{4}
\end{equation}
Notably, the glass transition is {\em completely suppressed} for strong enough coupling fluctuations, 
\begin{equation}
D_a > D_a^{*}\equiv \sqrt{4(p-2)^{p-2}/(p(p-1)^{p-1})}
\end{equation}
For $p=3$, $D_a^{*} = \sqrt{\frac{2}{3}} \approx 0.81$.

\begin{figure*}[t]
    \centering
    \includegraphics[width=\linewidth]{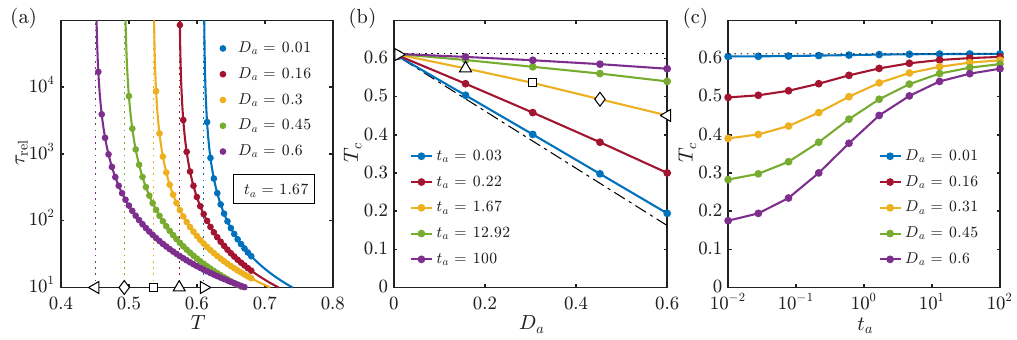}
    \caption{\textit{Phase diagram for $p$-spin model with dynamical couplings ---} (a) The decay time $\tau_{\rm decay}$ is plotted against $T$ for $t_a=1.67$ and varying $D_a$. We fit the data (dots) through a power law (solid lines) Eq.\,\ref{eq:Powerlaw} (solid lines) to obtain the critical temperature $T_c$ (vertical dotted lines) for each $D_a$, $t_a$. (b-c) The resulting phase diagram for this model: as the strength ($D_a$) and persistence ($t_a$) of the coupling fluctuations vary, we evaluate $T_c(D_a, t_a)$ as in (a) and how it varies between the two fully tractable cases: (i) white noise fluctuations ($t_a\rightarrow 0$, dot-dashed line) and (ii) frozen couplings ($t_a\rightarrow \infty$, dotted line).}
    \label{fig:phase_diagram}
\end{figure*}

In the glass phase, the correlation function $C(\tau)$ exhibits an infinite shoulder parametrized by the Edwards-Anderson parameter $C(\tau \rightarrow \infty) = q$. We determine $q$ for the current model to solve
\cite{Note1}
\begin{equation}
    T = \sqrt{\frac{p(p-1)q^{p-2}}{2}}(1-q) - \frac{pD_a}{4}.
\end{equation}
We stress that, unlike the thermal $p$-spin model, $q$ does not tend to $1$ as $T \rightarrow 0$ as the active fluctuations always generate a residual temperature. 

\textit{Infinitely persistent active fluctuations. ---} In the limit $t_a \to \infty$, the fluctuations in the couplings disappear and the system reduces to the standard thermal $p$-spin model at temperature $T$. This reflects the fact that energy injected on timescales much longer than the relaxation time $\tau_{\rm rel}$ is dynamically irrelevant: the fluid relaxes before the couplings evolve. In this regime, the activity effectively acts as an additional quenched disorder. We expect this regime to hold as long as $t_a \gg \tau_{\rm rel}$.

\textit{Active fluctuations with finite persistence.---} For finite $t_a$, the dynamics escape a direct mapping to that of a thermal $p$-spin model, thus describing a truly non-equilibrium model. Thus, we resort to solving Eq.\,(\ref{eq:stationaryeomCandR}) numerically (see \cite{Note1} for details of the numerics). The stationarity assumption allows us to explore the correlation function up $\tau = 10^7$ (i.e. over more than 10 decades). 

Motivated by mode-coupling theory, we numerically analyze the $p=3$ case, but expect our results to be qualitatively valid for $p\ge 3$. Figs.\,\ref{fig:shoulder_Da}(a-b) show that fluctuating couplings stabilize the liquid phase well below the equilibrium glass temperature $T_{g,{\rm eq}} = \sqrt{3/8}$ for finite activity strength $D_a$ and persistence time $t_a$. Remarkably, $D_a$ and $t_a$ play opposite roles: increasing $D_a$ fluidizes the system, while increasing $t_a$ counteracts this effect. Physically, $t_a$ cannot cool the system below $T$; it offsets the heating induced by $D_a$. 

The solutions $C(\tau)$ and $R(\tau)$ exhibit a clear timescale separation, with distinct effective FDRs for $\tau \ll t_a$ and $\tau \gg t_a$. Fig.\,\ref{fig:shoulder_Da}(c) shows the integrated response function $\chi(\tau) = -T\int_0^{\tau} d\tau' R(\tau')$ plotted against the correlation function. These curves display two linear regimes connected by a crossover, where the slopes define the effective temperatures.  For $\tau\ll t_a$, one finds $T_{\rm eff} = T$ as the fluctuations are effectively frozen, while for $\tau \gg t_a$, $T_{\rm eff}$ becomes non-trivial and depends on the activity parameters. The persistence time $t_a$ determines the onset of this late-time regime: it occurs immediately for very small $t_a$, but is delayed for larger $t_a$. 

In Fig.\,\ref{fig:T_eff_vs_ta}, we extract $T_{\rm eff}$ from the slope of Fig.\,\ref{fig:shoulder_Da}(c). The results agree with the analytic limits found above for $t_a \to 0$ and $t_a \to \infty$. Convergence slows down near the equilibrium glass transition because lowering $T$ increases $\tau_{\rm rel}$, requiring much larger $t_a$ to ensure $t_a \gg \tau_{\rm rel}$. Below $T_{g,{\rm eq}}$, however, no such decay to $T$ is observed: at these temperatures, the model cannot be mapped onto a thermal $p$-spin system even for $t_a \gg \tau_{\rm rel}$, as it would already be in the glass phase. Thus, the thermal equations do not approximate our dynamics in this regime (see \cite{Note1} for further discussion).

The presence of these two effective FDRs and an internally fixed timescale $t_a$ suggests a re-writing of the physical quantities in terms of ``slow" and ``fast" variables (e.g $C = C_s + C_f$), as employed in Refs.\,\cite{Berthier2013, Berthier2000a}. Integrating out the fast variables will result in an effective equation for the slow ones (identical to that in Ref.\,\cite{Berthier2013}):
\begin{subequations}
\label{eq:slowdynamicseffectiveeom}
\begin{align}
        \frac{\partial C_s}{\partial \tau} &= -(\mu - I_{\Sigma}) C_s(\tau) + I_R D_s(\tau) \nonumber \\ 
        &\quad + \int_0^{\infty}d\tau' \left[ \Sigma_s(\tau + \tau') C_s(\tau') + D_s(\tau + \tau') R_s(\tau')\right] \nonumber\\
        &\quad + \int_{0}^{\tau} d\tau' \Sigma_s(\tau - \tau') C_s(\tau')\\ 
        \frac{\partial R_s}{\partial \tau} &= -(\mu - I_{\Sigma}) R_s(\tau) + I_R \Sigma_s(\tau) \nonumber \\
	&\quad + \int_0^{\tau} d\tau' \Sigma_s(\tau - \tau')R_s(\tau') \\
        \mu &= T + \int_0^{\infty} d\tau'\left[ \Sigma_f(\tau') C_f(\tau') + D_f(\tau') R_f(\tau')\right] \nonumber \\
	&\quad + \int_0^{\infty} d\tau'\left[ \Sigma_s(\tau') C_s(\tau') + D_s(\tau') R_s(\tau')\right]
\end{align}
\end{subequations}
where $I_{\Sigma} = \int_0^{\infty} d\tau \Sigma_f(\tau)$ and $I_R = \int_0^{\infty} d\tau R_f(\tau)$. 

Combining the late-time effective FDRs with the slow dynamics yields a ``closed'' equation for $C_s(\tau)$, formally equivalent to a thermal $p$-spin model with activity renormalizing microscopic coefficients. This suggests that the glass transition survives and retains its essential features \cite{Berthier2013}, which we confirm numerically. To probe it, we compute the relaxation time
\begin{equation}
    \tau_{\rm rel} = \int_0^\infty d\tau\, C(\tau).
\end{equation}
In the thermal $p$-spin model, $\tau_{\rm rel}$ diverges {\em algebraically} with exponent $\gamma = 1.765$ as we approach the glass transition \cite{Berthier2011}. As shown in Fig.\,\ref{fig:phase_diagram}, we find the same scaling form close to the critical temperature,
\begin{equation}\label{eq:Powerlaw}
    \tau_{\rm rel} \sim (T-T_g)^{-\gamma},
\end{equation}
where the critical exponent $\gamma$ and the glass transition temperature $T_g(D_a,t_a)$ are found through a fitting procedure [see Fig.\,\ref{fig:phase_diagram}(b)-(c)]. For the thermal $p$-spin model, this is known to be $\gamma = 1.765$ \cite{Berthier2011}. We find that $\gamma$ is a non-universal exponent which depends explicitly on the activity parameters ranging between $\gamma  \in [1.72, 3.1]$ (see \cite{Note1} for a more detailed discussion).

\begin{figure}
\centering
\includegraphics[width=0.9\linewidth]{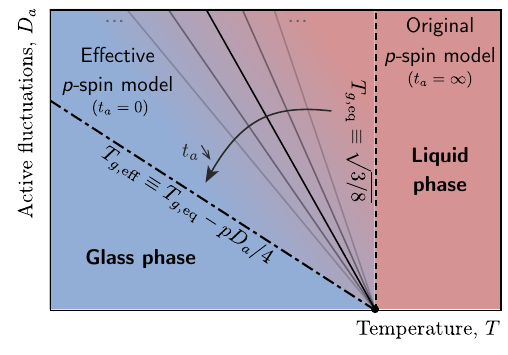}
\caption{\textit{Phase diagram ---} Schematic phase diagram of the $p$-spin model supplemented by active fluctuating couplings characterized by a strength $D_a$ and a persistence time $t_a$. We show that the glass transition in this model can be controlled by two parameters $(D_a,t_a)$; $T_g(D_a,t_a)$ varies smoothly between the original thermal $p$-spin model with $T_{g,{\rm eq}}$ when $t_a \to \infty$ and an effective thermal $p$-spin model with $T_{g,{\rm eq}}-pD_a/4$ when $t_a \to 0$. In particular, the glass transition can be partially suppressed by increasing $D_a$ or lowering $t_a$. For large enough active fluctuations, complete suppression of the glass transition can be observed as $T_g(D_a,t_a)$ becomes negative.}
\label{fig:phase_diagram_schematic}
\end{figure}

Fig.\,\ref{fig:phase_diagram_schematic} succinctly summarizes our results. Increasing the activity strength $D_a$ shifts the transition to lower temperatures and fluidizes the system, while larger persistence times $t_a$ promotes kinetic arrest. The critical temperature varies smoothly with $t_a$, interpolating between $T_{g,{\rm eq}}-pD_a/4$ and $T_{g,{\rm eq}}$; this dependence is shown explicitly in Fig.~\ref{fig:phase_diagram}(c). At large $D_a$, the value of $t_a$ becomes especially important in setting $T_g$.

\textit{Conclusions \& outlook.---} We have demonstrated how dynamic couplings can partially or even fully suppress the glass transition temperature in a non-equilibrium $p$-spin model. We show that the strength and persistence of interaction fluctuations have opposite effects: stronger fluctuations reduce the glass transition temperature, whereas more persistent fluctuations alleviates this effect and induces an earlier onset of the glass phase. We demonstrate this by extracting the critical temperature from Eq.\,(\ref{eq:Powerlaw}) as a function of the two parameters characterizing the fluctuations, $T_c(D_a, t_a)$.

Our results illustrate how fluctuating interactions drive transitions between glassy and fluid states. Interestingly, the tunability of the glass transition temperature $T_g$ via the fluctuation strength $D_a$ and persistence time $t_a$ provides a possible microscopic mechanism for dense living systems to regulate how far from the glass transition they operate, enabling reversible switching between fluid-like and rigid states during morphogenetic or physiological processes. We expect that our framework can be extended to mixed spin models and models where the correlations between the fluctuating couplings follow a power-law decay (lacking any inherent timescale $t_a$), for which classification of the onset of glassy dynamics remains an intriguing open problem. More generally, we hope to extend this bottom-up approach, which directly connects microscopic dynamics to macroscopic transitions, to finite-dimensional systems exhibiting \textit{structural} glass transitions. One route towards this would be to extend the mode-coupling theory for active matter to systems with fluctuating interaction forces.

\begin{acknowledgments}
We thank Ludovic Berthier for interesting discussions, in particular regarding the numerical algorithms. We also thank Chiu Fan Lee for interesting discussions. ES would like to thank Andy Thomas for invaluable technical support. HA and ES were supported by a Roth Scholarship at Imperial College London.
\end{acknowledgments}


%

\end{document}


\title{Supplemental Material for ``Controlling the Glass Transition through Active Fluctuating Interactions"}

\author{Emir Sezik}
\email{emir.sezik19@imperial.ac.uk}
\affiliation{Department of Mathematics, Imperial College London, South Kensington, London SW7 2AZ, United Kingdom}

\author{Henry Alston}
\email{henry.alston@phys.ens.fr}
\affiliation{Department of Mathematics, Imperial College London, South Kensington, London SW7 2AZ, United Kingdom}
\affiliation{Laboratoire de Physique, \'Ecole Normale Sup\'erieure, CNRS, PSL Universit\'e, Sorbonne Universit\'e, Universit\'e de Paris, 75005 Paris, France}

\author{Thibault Bertrand}
\email{t.bertrand@imperial.ac.uk}
\affiliation{Department of Mathematics, Imperial College London, South Kensington, London SW7 2AZ, United Kingdom}

\date{\today}

\maketitle

\hrule
\vspace{-3\baselineskip} 
\tableofcontents
\vspace{1em}
\hrule

\section{Nonequilibrium $p$-spin models with fluctuating couplings}

In the main text, we seek to understand how the introduction of active fluctuating couplings reshapes the glass transition in a spherical $p$-spin model. To do so, we study the glassy dynamics of the mean-field spherical $p$-spin model and add ``activity" to it in 2 different ways. Each scenario represents two different ways activity can enter into the equations of motion. We consider both \textit{intrinsic} and \textit{extrinsic} fluctuations on the couplings. In each case, we derive exact equations of motion for the correlation and response functions. Because the system is mean-field, i.e. every spin interact with every other spin, these two types of fluctuations will give the same effective equations of motion.  

\subsection{Intrinsic fluctuations}
We first consider intrinsic fluctuations in which the coupling constant between each pair of spins fluctuates \textit{independently}. The model is governed by the following Langevin equation
\begin{subequations}
    \label{eq:modelintrinsic}
    \begin{gather}
    \sigmadot_{i_1}(t) = - \orderedsum \Jp \sigmas -\orderedsum \etap \sigmas - \mu(t) \sigma_{i_1}(t) + \xi_{i_1}(t) \\
    \mathcal{P}\left[\Jp \right] \propto \exp\left[ -\frac{N^{p-1}}{ p!}\orderedsum (\Jp)^2 \right]\\
    \mathcal{P}\left[ \etap \right] \propto \exp \left[-\frac{N^{p-1}}{ p! D_A} \orderedsump \int dt \int dt'  \etap \cdot F^{-1}(t-t') \cdot \etapp \right] \\
    \langle \xi_i(t) \rangle = 0, \hspace{0.5cm} \langle \xi_i(t) \xi_j(t') \rangle = 2T\delta_{ij}\delta(t-t')
\end{gather}
\end{subequations}
where $\mu(t)$ enforces the spherical constraint $\sum_{i = 1}^N\sigma_i^2(t) = N$, $\etap$ and $\Jp$ are fully symmetric in their respective indices and $F^{-1}(t-t')$ is a linear operator satisfying
\begin{equation}
   \int dt' F^{-1}(t-t') F(t'-t'') = \delta(t-t'') 
\end{equation}
Now, we shall derive the effective equations of motion. We first rewrite the Langevin dynamics using a Martin-Siggia-Rose-Janssen-de Dominicis (MSRJD) path integral:
\begin{equation}
    Z_{J,\eta}[h] = \int \mathcal{D} [\sigma, \sigma] \exp\left[ \sums \int dt \sigma_i(t) h_i(t) - S[\sigma, \sigmahat]  \right]
\end{equation}
where the subscripts $J$ and $\eta$ indicate that the generating functional still depends on those variables. The action is given by 

\begin{align}
S[\sigma,\hat{\sigma}] = \sum_{i_1 = 1}^{N} \int dt\hat{\sigma}_{i_1}(t) \left[ \right. \dot{\sigma}_{i_1}(t) - \mu(t) \sigma_{i_1} (t) & + \sum_{i_2 < \dots < i_p}^{N} J^{(p)}_{i_1, i_2, \dots ,i_p} \sigma_{i_2}(t) \dots \sigma_{i_{p}}(t) \nonumber\\
    & + \left. \sum_{i_2 < \dots < i_p}^{N} \eta^{(p)}_{i_1, i_2, \dots ,i_p} \sigma_{i_2}(t) \dots \sigma_{i_{p}}(t) -T \hat{\sigma}_{i_1}(t)\right]
\label{eq:actionintrinsic}
\end{align}
There are two quantities we need to average over: $\Jp$ and $\etap$. Because these quantities are independent and they are not coupled in the Langevin equation, one can take the average independently. Specifically, we need to calculate the quantity:
\begin{align}
        \deltaav  \equiv\int \mathcal{D}J \mathcal{P}[J] \exp&\left( \sum_{i_1 = 1}^{N} \int dt \sigmahat_{i_1}(t)  \sum_{i_2 < \dots < i_p}^{N} \Jp \sigma_{i_2}(t) \dots \sigma_{i_{p}}(t) \right) \times \nonumber \\
        & \int \mathcal{D}[\eta] \mathcal{P}[\eta] \exp \left(  \sum_{i_1 = 1}^{N} \int dt \sigmahat_{i_1}(t) \sum_{i_2 < \dots < i_p}^{N} \etap \sigma_{i_2}(t) \dots \sigma_{i_{p}}(t) \right)
\end{align}
Using the now well established methods of Refs.\,\cite{Kirkpatrick1987a, Kirkpatrick1987b, Cugliandolo2003, Castellani2005}, we can calculate the averages. First, we reorder the summation:
\begin{align}
    \sum_{i_1}^{N}\sum_{i_2 < \dots < i_p}^{N} &\Jp \sigmahats  \nonumber\\
     & = \orderedsump \Jp \left[ \sigmahatssym \right] 
\end{align}
where we have used the symmetry of the coupling constant $\Jp$. This can be similarly done for $\etap$. Performing the integrals and collecting the resulting terms, we are left with 
\begin{align}
\label{eq:averageddelta}
        \deltaav = \exp \left[ \int dt\int dt'  \frac{1 + D_A F(t-t')}{4 N^{p-1}}\Big( p\sighatsighat \sigsig^{p-1} +p(p-1) \sighatsig \sigsighat \sigsig^{p-2} \Big)\right]
\end{align}
where we have defined 
\begin{subequations}
    \begin{align}
        \sigsig & \equiv \sums \sigma_i(t)\sigma_i(t'), \hspace{0.5cm} \sighatsighat \equiv \sums \sigmahat_i(t)\sigmahat_i(t')\\
        \sigsighat &\equiv \sums \sigma_i(t)\sigmahat_i(t'), \hspace{0.5cm} \sighatsig \equiv \sums \sigmahat_i(t)\sigma_i(t') 
    \end{align}
\end{subequations}
We shall now insert the following order parameters 
\begin{subequations}
    \begin{align}
        Q_1(t,t') & \equiv \frac{1}{N} \sighatsighat, \hspace{0.5cm} Q_2(t,t') \equiv \frac{1}{N} \sigsig\\
         Q_3(t,t') & \equiv \frac{1}{N} \sighatsig, \hspace{0.5cm} Q_4(t,t') \equiv \frac{1}{N} \sigsighat\\
    \end{align}
\end{subequations}
into the path integral through the following resolution of identity:
\begin{align}
    1 &= \int \mathcal{D}Q_1 \delta\left[ NQ_1(t,t') - \sighatsighat \right] = \int \mathcal{D}[Q_1,\lambda_1] \exp\left[ i \int dt \int dt' \lambda_1(t,t') (NQ_1(t,t') - \sighatsighat)\right]
\end{align}
where we have used the Fourier representation of a Dirac delta. Inserting these parameters, we obtain the following generating functional
\begin{align}
   \deltaav &= \int \mathcal{D}[Q_i, \lambda_i, \lambdahat_i] \exp\left[ NG + \Lambda \right]
\end{align}
where we have defined
\begin{subequations}
\label{eq:deltaop}
\begin{align}
    G &\equiv  \int dt \int dt' \left[\frac{1 + D_a F(t-t')}{4}(p Q_1 Q^{p-1}_2 + p(p-1)Q_3 Q_4 Q^{p-2}_2) \right]  + i\sum_{i = 1}^{4} \left[ \lambda_i(t,t') Q_i(t,t') \right] \\
    \Lambda &\equiv -i \int dt \int dt' \Big[ \lambda_1 \sighatsighat + \lambda_2 \sigsig + \lambda_3 \sighatsig +\lambda_4 \sigsighat \Big]
\end{align} 
\end{subequations}
In the thermodynamic limit, only the saddle point of this functional will contribute to the integral. The saddle-point equations read
\begin{subequations}
\begin{align}
    -i \lambda_1(t,t') &=  \frac{1 + D_a F(t-t')}{4} pQ^{p-1}_2(t,t')\\
    -i \lambda_2(t,t') &= \frac{1 + D_a F(t-t')}{4} p(p-1) \left( Q_1 Q_2^{p-2} + (p-2)Q_3 Q_4 Q_2^{p-2}\right) \\
    -i \lambda_3(t,t') &=  \frac{1 + D_a F(t-t')}{4} p(p-1) Q_4(t,t')Q^{p-2}_2(t,t')\\
    -i \lambda_4(t,t') &=  \frac{1 + D_a F(t-t')}{4} p(p-1) Q_3(t,t')Q^{p-2}_2(t,t')
\end{align} 
\end{subequations}
Furthermore, due to the self averaging in the large $N$ limit, we can replace quantities like $\frac{1}{N}\sigsig$ by their average. This allows us to identify 
\begin{align}
    Q_2(t,t') = C(t,t'), \hspace{0.25cm} Q_3(t,t') = R(t',t), \hspace{0.25cm} Q_4(t,t') = R(t,t')
\end{align}
Furthermore, by causality $Q_1(t,t') = 0$ and $Q_3(t,t')Q_4(t,t') = 0$.  Finally, we arrive at the following effective MSRJD functional
\begin{equation}
    \overline{Z}[h] = \exp\left[ \sum_{i = 1}^N\int dt \sigma_i(t) h_i(t) - S_{\rm eff} \right]
\end{equation}
with 
\begin{align}
    S_{\rm eff}[\sigma, \sigmahat] = \sums \int dt \sigmahat_i(t) \Bigg[ \sigmadot_i(t) - \mu(t) \sigma_i(t) &- \int dt' \frac{1 + D_a F(t-t')}{2} p(p-1) R(t,t') C^{p-2}(t,t') \sigma_i(t') \nonumber\\
    & - \int dt' \left( T\delta(t-t') + \frac{1 + D_a F(t-t')}{4} p C^{p-1}(t,t') \right) \sigmahat_i(t')\Bigg] 
    \label{eq:effaction}
\end{align}
This corresponds to the following effective Langevin equation of motion of a spin:
\begin{subequations}
\label{eq:effectivelangevineqintrinsic}
    \begin{gather}
        \sigmadot = -\mu(t) \sigma(t) + \int dt'\Sigma(t,t') \sigma(t') + \xi(t) \\
        \langle \xi(t) \xi(t') \rangle = 2T \delta(t-t') + D(t,t') 
    \end{gather}
\end{subequations}
where we have defined 
\begin{subequations}
\begin{align}
\Sigma(t,t') &= \frac{p(p-1)}{4}(1 + D_a F(t-t')) R(t,t')C^{p-2}(t,t')\\
D(t,t') &= \frac{p}{2}(1 + D_a F(t-t'))C^{p-1}(t,t')
\end{align}
\end{subequations}
and have suppressed the site index $i$ for ease of notation. 

\subsection{Extrinsic fluctuations}
In this section, we will investigate the effects of extrinsic fluctuations. In this case, extrinsic fluctuations refer to a scenario where the coupling constant between each spin fluctuates in the same way. The dynamics is here governed by the following Langevin equation
\begin{subequations}
\label{eq:modelextrinsic}
\begin{gather}
    \sigmadot_{i_1}(t) = - \orderedsum \Jp (1 + \eta(t))\sigmas - \mu(t) \sigma_{i_1}(t) + \xi_{i_1}(t) \\
    \mathcal{P}\left[\Jp \right] \propto \exp\left[ -\frac{N^{p-1}}{ p!}\orderedsum (\Jp)^2 \right]\\
    \mathcal{P}\left[ \eta \right] \propto \exp \left[-\frac{1}{D_A} \int dt \int dt' \eta(t) \cdot F^{-1}(t-t') \cdot \eta(t') \right] \\
    \langle \xi_i(t) \rangle = 0, \hspace{0.5cm} \langle \xi_i(t) \xi_j(t') \rangle = 2T\delta_{ij}\delta(t-t')
\end{gather}
\end{subequations}
where we have used the same notation as in the previous section.

To find the effective Langevin equation associated with this model, we cast Eq. (\ref{eq:modelintrinsic}) as an MSRJD path integral:
\begin{equation}
\label{eq:action}
    Z_{J,\eta}[h] = \int \mathcal{D} [\sigma, \sigma] \exp\left[ \sums \int dt \sigma_i(t) h_i(t) - S[\sigma, \sigmahat]  \right]
\end{equation}
with
\begin{align}
\label{eq:actionextrinsic}
        S[\sigma,\sigmahat] &= \sum_{i_1=1}^N \int dt\sigmahat_{i_1}(t) \left[ \sigmadot_{i_1}(t) - \mu(t) \sigma_{i_1} (t) + \sum_{i_2 < \dots < i_p}^{N} \Jp \big[1+ \eta(t)\big] \sigma_{i_2}(t) \dots \sigma_{i_{p}}(t) - T \sigmahat_{i_1}(t)  \right] 
\end{align}
In this case, the relevant quantity to average is
\begin{align}
\label{eq:deltaavextrinsic}
    \deltaav  \equiv\int \mathcal{D}[J,\eta] \mathcal{P}[J] \mathcal{P}[\eta] \exp\left( \sum_{i_1 = 1}^{N} \int dt \sigmahat_{i_1}(t)  \sum_{i_2 < \dots < i_p}^{N} \Jp 
   \big[1+\eta(t)\big]\sigma_{i_2}(t) \dots \sigma_{i_{p}}(t) \right) 
\end{align}
Unlike the previous case, the activity and quenched disorder are not decoupled. Therefore, one needs extra care to carry out this average. Again, we first reorder the summation:
\begin{align}
    \sum_{i_1}^{N}&\sum_{i_2 < \dots < i_p}^{N} \Jp \big[1+\eta(t)\big]\sigmahats  \nonumber\\
     & = \orderedsump \Jp \big[1+\eta(t)\big] \left[ \sigmahatssym \right] 
\end{align}
To ease the notation, we define 
\begin{equation}
    V_{i_1 i_2  \cdots i_p}(t) \equiv  \sigmahatssym 
\end{equation}
One can check that this tensor is fully symmetric in its indices. We first compute the $\eta$ integral in Eq. (\ref{eq:deltaavextrinsic}) which gives
\begin{align}
    \deltaav  \equiv\int \mathcal{D}[J] \mathcal{P}[J]  \exp\Bigg(  \orderedsump  &\int dt \Jp V_{i_1 i_2 \cdots i_p} (t)   \nonumber \\
    												&+ \orderedsump \sum_{j_1 < j_2 < \cdots < j_p}^{N} \Jp J^{(p)}_{j_1 j_2 \cdots j_p} X_{i_1, i_2, \cdots i_p, j_1, \cdots j_p}\Bigg) 
\end{align}
where we have defined 
\begin{equation}
    X_{i_1, i_2, \cdots i_p, j_1, \cdots j_p} = \frac{D_a}{4} \int dt \int dt' V_{i_1 i_2 \cdots i_p}(t) F(t-t') V_{j_1 j_2 \cdots j_p}(t') 
\end{equation}
We can now take the average over the quenched disorder $\Jp$:
\begin{align}
        \deltaav  \equiv\int \mathcal{D}[J] \  \exp\Bigg( & \orderedsump  \int dt \Jp V_{i_1 i_2 \cdots i_p} (t) \\
        &-\frac{N^{p-1}}{p!} \orderedsump \sum_{j_1 < j_2 < \cdots < j_p}^{N} \Jp \left[ \delta^{i_1, \cdots i_p}_{j_1, \cdots ,j_p} - \frac{p!}{N^{p-1}}X_{i_1, i_2, \cdots i_p, j_1, \cdots j_p}\right]  J^{(p)}_{j_1 j_2 \cdots j_p} \Bigg) \nonumber
        \label{eq:deltaavextrinsic2}
\end{align}
leading to 
\begin{align}
\deltaav = \exp&\left( -\frac{1}{2}\mathrm{Tr} \ln\left[  \boldsymbol{1} - \frac{p!}{N^{p-1}}\boldsymbol{X} \right] \right. \nonumber\\
		&\quad\quad+ \left.\orderedsump \sum_{j_1 < \cdots < j_p}^{N} \frac{p!}{4N^{p-1}} \int dt \int dt' V_{i_1 \cdots i_p}(t)  \left[\boldsymbol{1} - \frac{p!}{N^{p-1}} \boldsymbol{X} \right]^{-1}_{i_1, \cdots i_p, j_1, \cdots j_p} V_{j_1 \cdots j_p}(t')\right)
\end{align}
The final expression involves taking the inverse and the logarithm of the matrix $\boldsymbol{1} - \dfrac{p!}{N^{p-1}} \boldsymbol{X}$. To do this, we need to taylor expand $\boldsymbol{X}$. Due to the mean-field nature of the model, it will turn out that we shall only need at most the first order in the expansion.

Firstly, consider the logarithmic term. We can formally write it as
\begin{equation}
    \ln\left[ \boldsymbol{1} - \frac{p!}{N^{p-1} \boldsymbol{X}}\right] = -\sum_{n = 1}^{\infty} \frac{(p!)^n}{nN^{n(p-1)}}\boldsymbol{X}^n 
\end{equation}
On its own, this term will not vanish. However, when we insert the order parameters $Q_i$ as we did in the last section, this will vanish for $n \geq 2$. To see this, take the $n = 2$ term:
\begin{subequations}
\begin{align}
    \mathrm{Tr}\left[ \boldsymbol{X}^2 \right] &= \frac{D_a^2}{16} \orderedsump \sum_{j_1 < \cdots < j_p}^{N} \int \prod_{n = 1}^{4} dt_n F(t_1 - t_2) F(t_3 - t_4)  V_{i_1 \cdots i_p}(t_1) V_{i_1 \cdots i_p}(t_3)  V_{j_1 \cdots j_p}(t_2)  V_{j_1 \cdots j_p}(t_4) \\
    & = \frac{D_a^2}{16 p!^2}  \sum_{i_1, \cdots, i_p}^{N} \sum_{j_1, \cdots, j_p}^{N} \int \prod_{n = 1}^{4} dt_n F(t_1 - t_2) F(t_3 - t_4)  V_{i_1 \cdots i_p}(t_1) V_{i_1 \cdots i_p}(t_3)  V_{j_1 \cdots j_p}(t_2)  V_{j_1 \cdots j_p}(t_4) \\
    & = \frac{D_a^2}{16 p!^2} \int \prod_{n = 1}^{4} dt_n F(t_1 - t_2) F(t_3 - t_4) \left[ p\sighatsighat \sigsig^{p-1} +p(p-1) \sighatsig \sigsighat \sigsig^{p-2} \right](t_1,t_3) \times \nonumber\\
    &  \hspace{3cm}\left[ p\sighatsighat \sigsig^{p-1} +p(p-1) \sighatsig \sigsighat \sigsig^{p-2} \right](t_2,t_4)
\end{align}
\end{subequations}
Once the order parameters are inserted and the saddle-point approximation is taken, this term will vanish as there will always either be a factor of $Q_1(t,t')$ or $R(t,t')R(t',t)$ left which will nullify the contribution. Similar conclusions can be drawn for the higher order terms in the expansion.

Now, we consider the second term in Eq. (\ref{eq:deltaavextrinsic2}). Once again, $\left(\boldsymbol{1} - \dfrac{p!}{N^{p-1}} \boldsymbol{X}\right)^{-1}$ can be expanded as 
\begin{equation}
    \left(\boldsymbol{1} - \frac{p!}{N^{p-1}} \boldsymbol{X}\right)^{-1} = \sum_{n = 0}^{N} \left(\frac{p!}{N^{p-1}}\right)^n \boldsymbol{X}^n  
\end{equation}
Upon contraction with the $V$ tensors, the terms with $n \geq 1$ will vanish by the same line of reasoning. Therefore, the $\deltaav$ will be given by 
\begin{equation}
\label{eq:deltaaveraged2}
    \deltaav = \exp \left[ \int dt\int dt'  \frac{1 +\frac{D_a}{2} F(t-t')}{4 N^{p-1}}\left( p\sighatsighat \sigsig^{p-1} +p(p-1) \sighatsig \sigsighat \sigsig^{p-2} \right)\right]
\end{equation}
This equation is exactly Eq. (\ref{eq:averageddelta}) with a suitably rescaled $D_a$. Therefore, the effective Langevin equation for this model reads:
\begin{subequations}
\label{eq:effectivelangevineqextrinsic}
    \begin{gather}
        \sigmadot = -\mu(t) \sigma(t) + \int dt'\Sigma(t,t') \sigma(t') + \xi(t) \\
        \langle \xi(t) \xi(t') \rangle = 2T \delta(t-t') + D(t,t') 
    \end{gather}
\end{subequations}
where
\begin{subequations}
\begin{align}
\Sigma(t,t') &= \frac{p(p-1)}{2}\left(1 + \frac{D_a}{2} F(t-t')\right) R(t,t')C^{p-2}(t,t'), \\
D(t,t') &= \frac{p}{2}\left(1 + \frac{D_a}{2} F(t-t')\right)C^{p-1}(t,t').
\end{align}
\end{subequations}
It is only because the model is mean-field that we recover the same effective equation of motion as the intrinsic fluctuations. This result is quite non-trivial and we don't expect the result to hold for non mean-field models.

In the coming sections, we shall focus on analysing Eq. (\ref{eq:effectivelangevineqintrinsic}). However, one should bear in mind that whatever we say about the intrinsic fluctuations will also hold true in the case of extrinsic fluctuations [specifically model defined by Eq. (\ref{eq:effectivelangevineqextrinsic})] as well.

\section{Full equations for $C$ and $R$ from EOM}

Eqs.s (\ref{eq:effectivelangevineqintrinsic}) and (\ref{eq:effectivelangevineqextrinsic}) for a given spin involve the response and correlation functions. To get a closed set of equations that we can attempt to solve, we shall derive equations of motion for the correlation and the response function. 

First, recall that the effective Langevin equation for a single spin is given by 
\begin{equation}
\label{eq:effectivelangevin}
    \sigmadot_i(t) = -\mu(t) \sigma_i(t) + \int dt'\Sigma(t,t') \sigma_i(t') + \xi_i(t)
\end{equation}
where $\langle \xi_i(t) \xi_j(t') \rangle = 2T \delta(t-t') + D(t,t')$, $\Sigma(t,t') = \frac{p(p-1)}{2} C^{p-2}(t,t') R(t,t')(1+D_a F(t-t'))$ and $D(t,t') = \frac{p}{2} C^{p-1}(t,t')(1+D_a F(t-t'))$. Remembering that the correlation function is defined by $C(t,t') = \frac{1}{N}\sum_{i = 1}\langle \sigma_i(t) \sigma_i(t') \rangle$, we can multiply Eq. (\ref{eq:effectivelangevin}) by $\sigma_i(t')$ and sum over $i$ to get
\begin{equation}
    \frac{\partial C(t,t')}{\partial t} = -\mu(t) C(t,t') + \int^{t}_{-\infty} dt'' \Sigma(t,t'') C(t'',t') + \frac{1}{N}\sum_{i = 1}^{N} \langle \xi_i(t) \sigma_i(t') \rangle
\end{equation}
One can then show that \cite{Castellani2005}
\begin{equation}
\label{eq:equationforc}
    \frac{1}{N}\sum_{i = 1}^{N} \langle \xi_i(t) \sigma_i(t') \rangle = 2T R(t',t) + \int_{-\infty}^{t'} dt' D(t,t'') R(t',t'')
\end{equation}
We can also, derive an equation for the response function by remembering its definition $R(t,t') =  \dfrac{1}{N} \sum_{i = 1}^{N} \left \langle  \dfrac{\delta \sigma_i(t)}{\delta \xi_i(t')} \right \rangle$:
\begin{equation}
    R(t,t') = -\mu(t) R(t,t') + \int_{t'}^{t}  dt''\Sigma(t,t'') R(t'',t') + \delta(t-t')
\end{equation}
Note that causality imposes that $R(t,t') = 0$ if $t<t'$. There is a subtlety regarding $t = t'$. One can show that Dirac delta implies:
\begin{subequations}
\label{eq:Ratorigin}
\begin{gather}
    \lim_{t' \rightarrow t^{-}} R(t,t') \equiv R(t,t^{-}) = 1 \\
     \lim_{t' \rightarrow t^{+}} R(t,t') \equiv R(t,t^{+}) = 0
\end{gather}
\end{subequations}
To close the set of equations, we need to self-consistently fix $\mu(t)$. This Lagrange multiplier is here to ensure the spherical constraint $\dfrac{1}{N} \sum_{i = 1}^{N} \langle \sigma_i(t) \sigma_i(t')\rangle = C(t,t) = 1$. This condition implies that \cite{Cugliandolo2003}
\begin{equation}
   \left. \frac{\partial C(t,t')}{\partial t} \right|_{t' = t^{-}} + \left. \frac{\partial C(t,t')}{\partial t'} \right|_{t' = t^{-}} = 0
\end{equation}
This, combined with Eqs. (\ref{eq:equationforc}) and (\ref{eq:Ratorigin}) give us
\begin{subequations}
\label{eq:CandRgeneralnoic}
    \begin{align}
        \frac{\partial C(t,t')}{\partial t} &= -\mu(t) C(t,t') + \int_{-\infty}^{t} dt'' \Sigma(t,t'') C(t'',t') + \int_{-\infty}^{t'} dt'' D(t,t'')R(t',t'') + 2TR(t',t) \\
        \frac{\partial R(t,t')}{\partial t} &= -\mu(t) R(t,t') + \int_{t'}^{t} dt'' \Sigma(t,t'')R(t'',t') + \delta(t-t') \\
        \mu(t) &= T + \int_{-\infty}^{t} dt'\left[\Sigma(t,t') C(t,t') + D(t,t') R(t,t') \right]
    \end{align}
\end{subequations}
Throughout, we have implicitly assumed that the system was initialised at $t_0 \rightarrow -\infty$. We can also initialise the system at $t_0 = 0$ with random initial conditions \cite{Barrat1997, Berthier2001, Kim2001} for $\sigma_i$ and $\etap$ at its steady state. To implement the choice of initial conditions, we need to average over them in Eq. (\ref{eq:action}). However, because random initial conditions correspond to a uniform distribution of the $\sigma_i(0)$, this choice simply amounts to replacing $-\infty$ with 0 in the above equations. Therefore, within this choice of initialisation, we have
\begin{subequations}
\label{eq:CandRgeneralRIC}
    \begin{align}
        \frac{\partial C(t,t')}{\partial t} &= -\mu(t) C(t,t') + \int_{0}^{t} dt'' \Sigma(t,t'') C(t'',t') + \int_{0}^{t'} dt'' D(t,t'')R(t',t'') + 2TR(t',t) \\
        \frac{\partial R(t,t')}{\partial t} &= -\mu(t) R(t,t') + \int_{t'}^{t} dt'' \Sigma(t,t'')R(t'',t') + \delta(t-t') \\
        \mu(t) &= T + \int_{0}^{t} dt'\left[\Sigma(t,t') C(t,t') + D(t,t') R(t,t') \right]
    \end{align}
\end{subequations}
A priori, there is no reason for the correlation and response functions to be stationary in time. Indeed, in the thermal $p$-spin model, there exists a critical temperature $T_g$ below which the dynamics take infinite time to be stationary \cite{Cugliandolo1993, Cugliandolo2003}. However, above this critical temperature and for sufficiently long waiting times $t'$, the dynamics are stationary in the $p$-spin model. Following \cite{Berthier2013, Berthier2000a}, we analyze the stationary equations. The stationary equations can be readily obtained by letting $C(t,t') = C(t-t')$ and $R(t-t')$ in Eq. (\ref{eq:CandRgeneralnoic}). These are given by 
\begin{subequations}
    \label{eq:CandRstationary}
    \begin{align}
         \frac{ \partial C(\tau)}{\partial \tau} &= -\mu C(\tau) + \int_0^{\tau} dt \tau' \Sigma(\tau - \tau') C(\tau') + \int_{0}^{\infty} d\tau' \left[ \Sigma(\tau + \tau') C(\tau') + D(\tau + \tau') R(\tau') \right] \\
         \frac{\partial R(\tau)}{\partial \tau} &= -\mu R(\tau) + \int_{0}^{\tau} d\tau' \Sigma(\tau - \tau') R(\tau') +\delta(\tau) \\
         \mu &= T + \int_{0}^{\infty} d\tau' \left[ \Sigma(\tau') C(\tau') + D(\tau') R(\tau') \right]
    \end{align}
\end{subequations}
To arrive at this conclusion, we have used the fact that $C(\tau)$ is a symmetric function and that a stationary regime exists. Specifically, we have assumed that $F(t-t')$ is a bounded function. One can obtain the same result from Eq. (\ref{eq:CandRgeneralRIC}) by sending $t$ and $t'$ to infinity while keeping $\tau = t-t'$ constant. This form is more amenable for numerical integration as only the time difference is relevant for the dynamics. 

\section{Delta-correlated fluctuations: Exact mapping to thermal $p$-spin at an effective temperature}

Up until now, we have kept the correlation function of the active noise $F(t-t')$ general. To understand intuitively its role on the dynamics, in this section we set it to a Dirac delta. One can easily show that Eq. (\ref{eq:CandRgeneralnoic}) then reduces to 
\begin{subequations}
\label{eq:CandRwhitenoise}
    \begin{align}
        \frac{\partial C(t,t')}{\partial t} &= -\mu(t) C(t,t') + \int_{-\infty}^{t} dt'' \Sigma_{th}(t,t'') C(t'',t') + \int_{-\infty}^{t'} dt'' D_{th}(t,t'')R(t',t'') + 2\left(T + \frac{p D_a}{4} \right)R(t',t) \\
        \frac{\partial R(t,t')}{\partial t} &= -\mu(t) R(t,t') + \int_{t'}^{t} dt'' \Sigma_{th}(t,t'')R(t'',t') + \delta(t-t') \\
        \mu(t) &= T + \frac{pD_a}{4} + \int_{0}^{\infty} dt'\left[\Sigma_{th}(t,t') C(t,t') + D_{th}(t,t') R(t,t') \right]
    \end{align}
\end{subequations}
where 
\begin{equation}
\Sigma_{\rm th}(t,t') = \frac{p(p-1)}{2} C^{p-2}(t,t') R(t,t')\quad \mathrm{and}\quad D_{\rm th}(t,t') = \frac{p}{2}C^{p-1}(t,t')
\end{equation} 
are the terms one gets when considering a thermal $p$-spin model. The structure of Eq. (\ref{eq:CandRwhitenoise}) suggests that the dynamics of this model is equivalent to a thermal $p$-spin model with an effective temperature $T_{\rm eff} = T + \dfrac{p D_a}{4}$. In this case, the addition of the activity effectively heats up the system. Furthermore, we also recover the fluctuation-dissipation relation (FDR) between the stationary quantities $C(\tau)$ and $R(\tau)$:
\begin{equation}
    \label{eq:effectiveFDt}
    R(\tau) = -\frac{\theta(\tau)}{T_{\rm eff}}\frac{\partial C(\tau)}{\partial \tau}
\end{equation}
where $\theta(\tau)$ is the Heaviside theta function. We use this relation to derive a closed equation for $C(\tau)$:
\begin{equation}
    \label{eq:whitenoiseC}
    \frac{\partial C(\tau)}{\partial \tau} = - T_{\rm eff}C(\tau) - \frac{p}{2T_{\rm eff}} \int_{0}^{\tau} d\tau' C^{p-1}(\tau - \tau')\frac{\partial C(\tau')}{\partial \tau'}
\end{equation}
This equation characterizes completely the stationary dynamics of the system and we will solve it numerically in a later section. After solving this equation, one can trivially find the response function using the FDR.  

For thermal $p$-spin models, there exists a critical temperature $T_g$ below which the system takes infinite time to equilibriate and the correlation plateaus to some value $q$. For pure spherical $p$-spin models, this temperature is given by 
\begin{equation}
    T_{g, {\rm eq}} =  \sqrt{\frac{p(p-2)^{p-2}}{2(p-1)^{p-1}}}
\end{equation}
In our case, this temperature gets shifted down due to the activity:
\begin{equation}
\label{eq:glasstransitionT}
    T_{g, {\rm eff}} =  \sqrt{\frac{p(p-2)^{p-2}}{2(p-1)^{p-1}}} - \frac{pD_a}{4}
\end{equation}
It is evident that if the strength of the activity exceeds the critical value $D_a^{*} = \sqrt{4(p-2)^{p-2}/(p(p-1)^{p-1})}$, the glass transition gets \textit{completely} suppressed. 

Below the critical temperature, for the thermal $p$-spin model, the Edward-Anderson parameter ($q \equiv C(\tau \rightarrow \infty)$) takes the form
\begin{equation}
    T = \sqrt{\frac{p(p-1)q^{p-2}}{2}}(1-q)
\end{equation}
In our case, the same picture will hold with a shifted temperature. Specifically, we will have the following relation between $q$ and $T$: 
\begin{equation}
    T = \sqrt{\frac{p(p-1)q^{p-2}}{2}}(1-q) - \frac{pD_a}{4}
\end{equation}
Unlike its thermal counterpart, in this limit, $q$ doesn't tend to $1$ as we lower the temperature to $0$. This is because there is always a residual temperature stemming from the activity.

\section{Exponentially correlated fluctuations with timescale $t_a$}

In this section, we shall consider an exponentially correlated noise. It is given by
\begin{equation}
    F(t-t') = \frac{1}{2t_a}\exp\left(-\frac{|t-t'|}{t_a}\right)
\end{equation}
where $t_a$ is the persistence time of the activity. Crucially, it doesn't depend on the dynamics of the system and is fixed by its own internal dynamics. This case is truly non-equilibrium as this model cannot be mapped to a thermal $p$-spin model. Therefore, it is not immediately obvious whether the glass transition can be completely suppressed, we expect in general the transition to persist but be of non-equilibrium nature similarly to the transition observed in \cite{Berthier2013}. 

Before, we argue for the existence of the glass transition, we consider two relevant limits. In the limit $t_a \rightarrow 0$, $F(t-t')$ formally approaches a Dirac delta function and we recover the case studied in the previous section. This limit corresponds to noise having zero persistence. From the previous section, we know that it corresponds to a thermal $p$-spin model with some effective temperature $T_{\rm eff}$. One other interesting limit corresponds to $t_a \rightarrow \infty$. In this limit, it takes an infinite time for the noise to fluctuate and therefore the active noise becomes, effectively, a quenched variable akin to $\Jp$. Therefore, we also recover a thermal $p$-spin model but now with a modified coupling strength and with temperature $T$.

\subsection{Existence of a glass phase}

In this subsection, we shall argue that for exponentially correlated active fluctuations the glass transition persists. By our definition, the persistence time of the activity is independent from the glassy dynamics. This will induce a strong timescale separation between the slow glassy dynamics and the fast dynamics due to the activity. Indeed, this separation of timescales is manifest  in our numerics and illustrated by an emergent fluctuation-dissipation relation with an effective temperature during late times. Due to this scale separation, we can separate any quantity into "slow" and "fast" variables (e.g $C = C_s + C_f$) similarly to what was done in \cite{Berthier2013, Berthier2000a}. Integrating out the fast variables will result in an equation for the slow variables identical to that presented in \cite{Berthier2013}:
\begin{subequations}
\label{eq:Equationsforslowdynamics}
    \begin{align}
        \frac{\partial C_s}{\partial \tau} &= -(\mu - I_{\Sigma}) C_s(\tau) + I_R D_s(\tau) + \int_{0}^{\tau} d\tau' \Sigma_s(\tau - \tau') C_s(\tau') + \int_0^{\infty}d\tau' \left[ \Sigma_s(\tau + \tau') C_s(\tau') + D_s(\tau + \tau') R_s(\tau')\right] \\
        \frac{\partial R_s}{\partial \tau} &= -(\mu - I_{\Sigma}) R_s(\tau) + I_R \Sigma_s(\tau) + \int_0^{\tau} d\tau' \Sigma_s(\tau - \tau')R_s(\tau') \\
        \mu &= T + \Omega + \int_0^{\infty} d\tau'\left[ \Sigma_s(\tau') C_s(\tau') + D_s(\tau') R_s(\tau')\right]
    \end{align}
\end{subequations}
where $I_{\Sigma} = \int_0^{\infty} d\tau \Sigma_f(\tau)$, $I_R = \int_0^{\infty} d\tau R_f(\tau)$ and $\Omega = \int_0^{\infty} d\tau'\left[ \Sigma_f(\tau') C_f(\tau') + D_f(\tau') R_f(\tau')\right]$. Combined with the emergent fluctuation dissipation relation, one can show that the resulting dynamics for the slow variable $C_s$ resembles that in Eq. (\ref{eq:whitenoiseC}) with activity renormalizing the microscopic coefficients. This was also found to be the case in Ref. \cite{Berthier2013} [see for instance Eq. (11)]. This suggests that our model does indeed display a non-equilibrium glass transition, whose nature is distinct from a thermal $p$-spin model due to the active fluctuations and the associated fast dynamics. These claims will be tested numerically in the next section where we show the existence of an emergent fluctuation-dissipation relation for different values of $t_a$ and a shoulder forming in the stationary correlation function for low enough temperatures which are standard observations pointing at the existence of a glass transition.

\subsection{Effective fluctuation-dissipation relation}

\begin{figure}[h!]
     \centering
     \includegraphics[width=\textwidth]{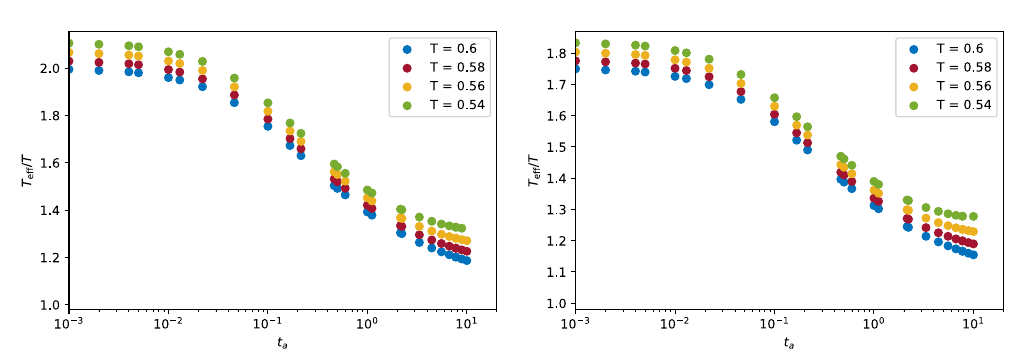}
         \caption{Late times effective temperature against the persistence of activity for (a) $D_a = 0.603$ and (b) $D_a = 0.8$. }
         \label{fig:Teffvsta}
\end{figure}

In this section, we further examine the violation of the fluctuation–dissipation relation (FDR) and the emergence of an effective temperature. As discussed in the main text, the persistence time of the activity leads to a separation of timescales, which in turn leads to two distinct FDRs depending on whether $\tau \ll t_a$ or $\tau \gg t_a$. For $\tau \ll t_a$, the couplings are effectively frozen, and the system is characterized by the physical temperature $T$. By contrast, for $\tau \gg t_a$, a nontrivial effective temperature $T_{\rm eff}$ emerges, which depends on the activity parameters. We showed in the main text that $T_{\rm eff} \to T$ as $t_a \to \infty$ when the system is above the equilibrium glass transition temperature. Below the glass transition, however, this convergence no longer occurs, as illustrated in Fig.\,\ref{fig:Teffvsta}. As expected, in the limit $t_a \to 0$, the effective temperature $T_{\rm eff}$ approaches $T + pD_a/4$. But in contrast to the high-temperature case, for large $t_a$ the effective temperature does \textit{not} relax to $T$. The reason, as argued in the main text, is that $t_a$ never becomes sufficiently large to exceed the intrinsic glassy relaxation timescales. In fact, the system undergoes a glass transition before such a regime can be reached, at which point the notion of an effective FDR ceases to be meaningful. Consequently, when $T < T_{g,\rm eq}$, there always exists a persistence time $t_a$ beyond which $T_{\rm eff}$ becomes ill-defined.

\subsection{Divergence of relaxation time}
In this section, we provide more details regarding the relaxation time. As discussed in the main text, this quantifies the shoulder formation and can be used to identify a glass transition. It is known that the relaxation time diverges as we approach the glass transition \cite{Berthier2011}. The relaxation time, $\tau_{\rm rel}$, can be defined as 
\begin{equation}
\tau_{\rm rel} = \int_0^{\infty} d\tau C(\tau).
\end{equation}
Plotting this quantity for a given $D_a$ and $t_a$ for different temperatures, we see that it increases algebraically as we lower the temperature. A similar behavior is observed in the case of the thermal $p$-spin model, where the relaxation time is known to diverge as
\begin{equation}
    \tau_{\rm rel} \sim \frac{1}{(T - T_g)^\gamma}
\end{equation}
We find that such a power law divergence still describes our data well, albeit with a different critical exponent $\gamma$. To deduce the values of $T_g$ and $\gamma$, we fit our numerically measured relaxation time $\tau_{\rm rel}$ to the following functional form:
\begin{equation}
    f(T; \gamma, T, A) = \frac{A}{(T - T_g)^\gamma}
\end{equation}
Since we expect this scaling to be true close to the critical point, we restrict the fitting range to values of the relaxation time $\tau_{\rm rel}$ greater than 100. 
\begin{figure}
     \centering
             \includegraphics[width=\textwidth]{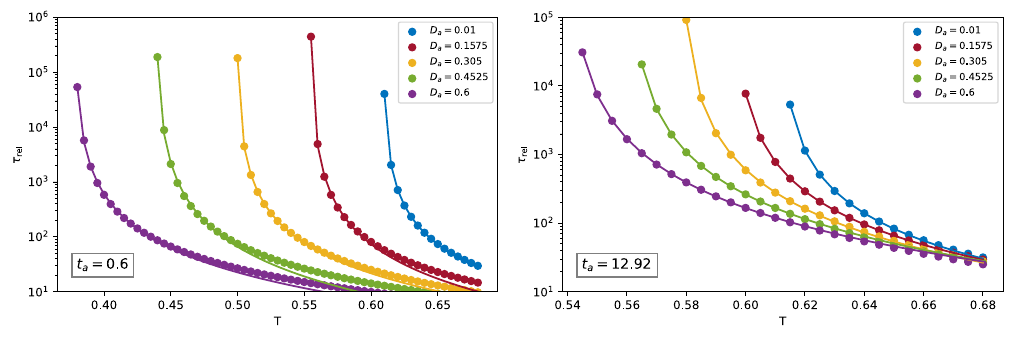}
         \caption{Decay time against temperature for (a) $t_a = 0.6$ and (b) $t_a = 12.92$. For large values of relaxation time, the divergence is governed by a power-law behaviour with an exponent $\gamma$.}
         \label{fig:Fittingexample}
\end{figure}
This procedure is illustrated in Fig.\,\ref{fig:Fittingexample}. We find a good agreement with our numerical data confirming the algebraic divergence of the relaxation time at the glass transition; the measured values of $T_g$ are discussed in the main text. 

\begin{figure}[h!]
    \centering
    \includegraphics[width=0.6\linewidth]{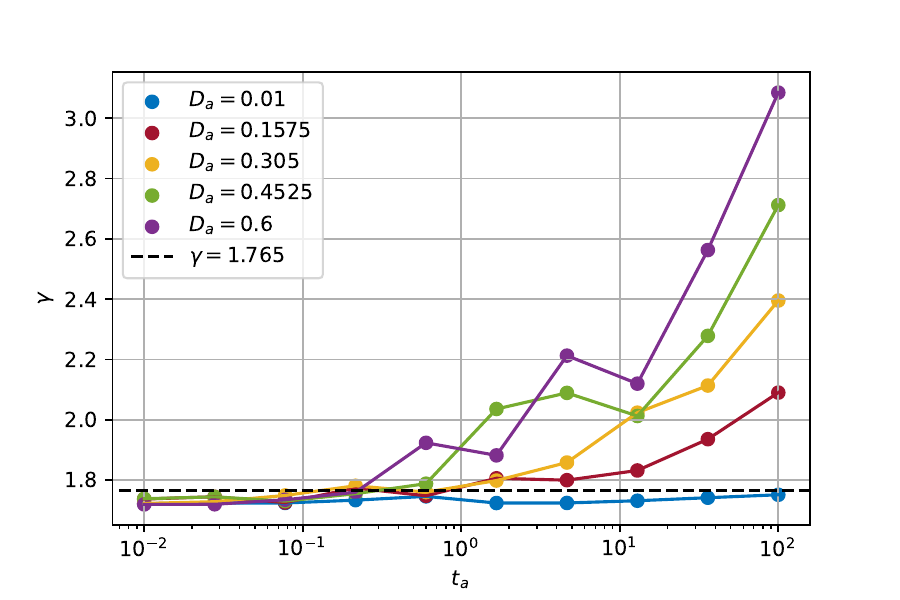}
    \caption{Values of the relaxation timescale divergence exponent $\gamma$ as a function of $t_a$ for different activity strength $D_a$. The dashed black line is the $\gamma$ value for the thermal $p$-spin model. In the equilibrium limit $D_a \rightarrow 0$ or $t_a \rightarrow 0$, we see that the measured exponent is close to the expected equilibrium value. Otherwise, the value of $\gamma$ depends on the activity parameters and it increases for larger $D_a$ and $t_a$ values. }
    \label{fig:gammavals}
\end{figure}

Through this fitting, we also infer the divergence exponent $\gamma$. We find that $\gamma$ is a non-universal exponent whose value depends on the activity parameters. This finding is illustrated in Fig.\,\ref{fig:gammavals}. The value of $\gamma$ tends to increase with the persistence time but the amount with which it increases depends on the activity strength.

\section{Details of numerical simulations}

In this section, we describe the numerical procedure used to solve Eq. (\ref{eq:CandRstationary}). Since Eq. (\ref{eq:CandRstationary}) is non-causal in time—requiring information at $\tau = \infty$ to evolve forward—we employ an iterative scheme. As the initial guess, we use the equilibrium solution with $D_a = 0$, i.e. solving Eq. (\ref{eq:whitenoiseC}) with $T_{\rm eff} = T$. The numerical method for solving both the equilibrium and active equations of motion is standard, but for completeness we briefly summarize it here.

\subsection{Equilibrium limit}
In the equilibrium limit, the FDR holds and Eq.\,(\ref{eq:CandRstationary}) reduce to Eq.\,(\ref{eq:whitenoiseC}). An efficient algorithm that solves equations of these type to many decades has been proposed in Ref.\,\cite{Fuchs1991}. We refer the reader to \cite{Flenner2005, Miyazaki2004} for a pedagogical discussion. We give a brief account of the algorithm for completeness. 

Our starting point is Eq.\,(\ref{eq:whitenoiseC}) with $T_{\rm eff} = T$. We first rescale time by temperature $\tau \rightarrow T \tau$ which leads to 
\begin{equation} \label{eq:dedimensionalisedeqequations}
    \dot{C}(\tau) = -C(\tau) - \frac{p}{2 T^2} \int_0^{\tau} d \tau' C^{p-1}(\tau - \tau') \dot{C}(\tau')
\end{equation}
It turns out this equation can be integrated exactly without the need for an iterative scheme due to its causal nature. Having set a time $T$ up to which we want to determine the correlation function, we first discretize the time into $N_T$ intervals of length $\delta$, such that $T = N_T \delta$ and ensure that $N_T = 2^{n_T}$ as working with powers of two is numerically more convenient. Here, we importantly assume that we know $C(\tau)$ at times $\tau_i = i \delta$ where the index $i$ ranges over $0 \leq i \leq N_T/2$ (i.e. the first-half of our time interval). We now aim to determine the value $C(\tau)$ takes at times $\tau_i = i\delta$ with  $N_T/2+1 \le i \le N_T$ (i.e. the second half of our interval). This can be done sequentially as the equations are causal. To do so, we discretize the integral at time $\tau_i = i \delta$ in the RHS of Eq. (\ref{eq:dedimensionalisedeqequations}) as follows
\begin{subequations}
    \begin{align}
        \int_0^{\tau_i} d \tau' m(\tau_i - \tau') \dot{C}(\tau') & = m(\tau_i - \tau_2) C(\tau_2) - m(\tau_i) - \int_{0}^{\tau_2} \frac{\partial m(\tau_i - \tau')}{\partial \tau'}C(\tau') + \int_{\tau_2}^{\tau_i} d \tau' m(\tau_i - \tau') \frac{\partial C(\tau')}{\partial \tau'} \\
        & = m(\tau_i - \tau_2) C(\tau_2) - m(\tau_i) - \sum_{j = 1}^{i_2} \int_{\tau_{j-1}}^{\tau_j} d\tau' \frac{\partial m(\tau_i - \tau')}{\partial \tau'}C(\tau') - \sum_{j = 1}^{i - i_2}\int_{\tau_{j-1}}^{\tau_j} d \tau'' m(\tau'') \frac{\partial C(\tau_i - \tau'')}{\partial \tau''}
    \end{align}
\end{subequations}
where we have performed the change of variable $\tau'' = \tau_i-\tau'$ in the last integral above and have defined $m(\tau) = \frac{p}{2T^2}C^{p-1}(\tau)$, $\tau_2 = \tau_i/2$ and by definition $i_2 = \lfloor i/2 \rfloor$. We here split the integral over two ranges as is commonly done because, within each range, the function whose derivative is taken varies slowly ensuring higher overall accuracy of the method. Note that there is some arbitrariness in the choice of the intervals; in splitting our time interval in half, we have followed previous literature \cite{Flenner2005, Miyazaki2004}. The integrals can be simplified by using the following approximation \cite{Fuchs1991}
\begin{equation} \label{eq:integralapprox}
    \int_{\tau_{j-1}}^{\tau_j} d\tau' \dot{A}(\tau') B(\tau') \approx \big(A(\tau_{j}) - A(\tau_{j-1})\big)\frac{1}{\delta} \int_{\tau_{j-1}}^{\tau_j} d\tau'  B(\tau') = \big(A(\tau_{j}) - A(\tau_{j-1})\big) dB_{j}
\end{equation}
where we have defined $dB_j = \frac{1}{\delta} \int_{\tau_{j-1}}^{\tau_j} d\tau'  B(\tau')$. With an appropriate integration scheme this approximation can be accurate to order $\delta^3$ \cite{Fuchs1991}. This is justified rigorously in Ref.\,\cite{Fuchs1991}. Specifically, we have chosen to approximate these integrals using Simpson's rule leading to 
\begin{equation} \label{eq:simpsonrule}
    dB_j = \frac{1}{12}\left( 5B_j + 8B_{j-1} - B_{j-2}\right)
\end{equation}
where we have used transparent notations and $B_j \equiv B(\tau_j)$. 

This improves the accuracy compared to applying the approximation in Eq. (\ref{eq:integralapprox}) directly to the original integral. Finally, discretizing the time derivative following Refs.\,\cite{Flenner2005, Miyazaki2004} leads to:
\begin{equation} \label{eq:derivativediscretisation}
   \dot{C}_i = \frac{1}{2 \delta} \left( 3 C_i - 4C_{i-1} + C_{i-2}\right) + \mathcal{O}(\delta^3) 
\end{equation}
allowing us to rewrite Eq. (\ref{eq:dedimensionalisedeqequations}) as 
\begin{equation} \label{eq:discretsiedeqforC}
    C_i = am_i + b_i
\end{equation}
where 
\begin{subequations}
    \begin{gather}
        a = \frac{1 - dC_1}{\frac{3}{2\delta} + 1 + dm_1} \\
        b_i = \frac{-m_{i - i_2} C_{i_2} + m_{i-1}dC_1 + C_{i-1}dm_1 + \frac{2}{\delta}C_{i-1}  - \frac{1}{2\delta}C_{i-2} 
  + \sum_{j = 2}^{i_2}(m_{i - j}- m_{i - j+1})dC_j + \sum_{j = 2}^{i - i_2}(C_{i - j} - C_{i - j+1})dm_j}{{\frac{3}{2\delta} + 1 + dm_1}}
    \end{gather}
\end{subequations}
where the integral bond notation $dC_i$ is defined above. Note that for index $i=1$, we replace Simpson's rule by the trapezoidal rule. The nonlinear discretized equation for $C_i$ can be easily solved. Once Eq.\,(\ref{eq:discretsiedeqforC}) is solved for all indices $(N_t+1)/2 \leq i \leq N_t$, we save the values of $C_i$ and apply a decimation transformation for the next iteration: we discard every other point in effect doubling the grid spacing $\delta \rightarrow 2\delta$. Under this decimation, the quantities transform as 
\begin{subequations}
    \begin{align}
        C_{2i} &\rightarrow C_i, &\hspace{0.2cm} 0\leq i \leq \frac{N_t}{2} \\
        m_{2i} &\rightarrow m_i, &\hspace{0.2cm} 0\leq i \leq \frac{N_t}{2} \\
        \frac{1}{2}(dC_{2i} + dC_{2i - 1}) &\rightarrow dC_i, &\hspace{0.2cm} 0\leq i \leq \frac{N_t}{4} \\
        \frac{1}{2}(dm_{2i} + dm_{2i - 1}) &\rightarrow dm_i, &\hspace{0.2cm} 0\leq i \leq \frac{N_t}{4} \\
        \frac{1}{6}(dC_{2i} + 4 dC_{2i - 1} + dC_{2i-2}) &\rightarrow dC_i, &\hspace{0.2cm} \frac{Nt}{4}+1\leq i \leq \frac{N_t}{2} \\
        \frac{1}{6}(dm_{2i} + 4 dm_{2i - 1} + dm_{2i-2}) &\rightarrow dm_i, &\hspace{0.2cm} \frac{Nt}{4}+1\leq i \leq \frac{N_t}{2}
    \end{align}
\end{subequations}
We now return to the original starting point, but with knowledge of the solution over a time interval twice as long. Specifically, we have a time grid of size $N_t$ with spacing $2\delta$, and the function is known for $0 \leq i \leq N_t/2$; our goal is to extend it to $N_t/2+1 \leq i \leq N_t$. The same equations and methods can then be applied iteratively. Repeating this procedure allows us to reach arbitrarily long times, and the decimation scheme enables access to many decades efficiently. 

To implement the algorithm, we require the values of $C(\tau)$ over an initial range $0 \leq \tau \leq N_t \delta/2$. For sufficiently small $\delta$, these can be obtained by propagating the initial condition. Expanding $C(\tau) = 1 + A\tau + \mathcal{O}(\tau^2)$ and substituting into Eq.,(\ref{eq:discretsiedeqforC}) yields $A = -1$. To ensure the validity of this linear approximation, we choose $\delta = 10^{-10}$ and $N_t = 1024$. With these parameters, the solution of Eq.,(\ref{eq:discretsiedeqforC}) can be efficiently computed up to times $\tau \sim 10^{16}$.

\subsection{Solution to non-causal equations}

We now turn to the solution of the full non-causal equations of motion. It is advantageous to work with slowly varying functions, since the approximation in Eq.\,(\ref{eq:integralapprox}) is more accurate in this case. To this end, we introduce the integrated response function
\begin{equation}
\chi(t,t') = -T\int_{t'}^t ds\,R(t,s)
\end{equation}
which varies more slowly than the bare response function \cite{Kim2001}. With this definition, and after rescaling time as $t \to Tt$, Eq.\,(\ref{eq:CandRgeneralnoic}) becomes
\begin{subequations}
    \begin{align}
        \frac{\partial C(t,t')}{\partial t} &= -\mu(t) C(t,t') +  \int_{t'}^{t} ds\, \sigma(t,s) \frac{\partial\chi(t,s)}{\partial s}C(s,t') + \int_{-\infty}^{t'}ds\, m(t,s) \frac{\partial \chi(t',s)}{\partial s} +\int_{-\infty}^{t'} ds\,\sigma(t,s) \frac{\partial \chi(t,s)}{\partial s} C(s,t') \\
        \frac{\partial \chi(t,t')}{\partial t} &= -1 -\mu(t) \chi(t,t') + \int_{t'}^{t} ds\,\sigma(t,s) \frac{\partial \chi(t,s)}{\partial s} \chi(s,t') \\
        \mu(t) &= 1 + p\int_0^t ds\,m(t,s) \frac{\partial \chi(t,s)}{\partial s}
    \end{align}
\end{subequations}
where we have defined $\sigma(t,s) = \frac{p(p-1)}{2T^2}C^{p-2}(t,s)\left(1 + D_a F \left(\frac{t - s}{T} \right)\right)$ and $m(t,s) = \frac{p}{2T^2}C^{p-1}(t,s)\left(1 + D_a F \left(\frac{t - s}{T} \right)\right)$. When the system is initialised in the infinite past, the dynamics eventually becomes stationary and we can let $C(t,t') = C(\tau)$ and $\chi(t,t') = \chi(\tau)$ where $\tau = t-t'$. Doing so, the equations above reduce to
\begin{subequations}  \label{eq:Candchinoic}
    \begin{align}
        \frac{\partial C(\tau)}{\partial \tau} &= -\mu C(\tau) +  \int_{0}^{\tau} d\tau' \sigma(\tau-\tau') \frac{\partial\chi(\tau-\tau')}{\partial \tau'}C(\tau') - \int_{0}^{\infty}d\tau' \left[ m(\tau + \tau') \frac{\partial \chi(\tau')}{\partial \tau'} +\sigma(\tau + \tau') \frac{\partial \chi(\tau + \tau')}{\partial \tau'} C(\tau') \right]\\
        \frac{\partial \chi(\tau)}{\partial \tau} &= -1 -\mu \chi(\tau) + \int_{0}^{\tau} d\tau' \sigma(\tau - \tau') \frac{\partial \chi(\tau - \tau')}{\partial \tau'} \chi(\tau') \\
        \mu(t) &= 1 - p\int_0^\infty d\tau' m(\tau') \frac{\partial \chi(\tau')}{\partial \tau'}
    \end{align}
\end{subequations}

In equilibrium, the FDR implies that $\chi(\tau) = C(\tau) - 1$. Unlike the causal equation in Eq.\,(\ref{eq:dedimensionalisedeqequations}), the full equations cannot be solved exactly, so we employ an iterative scheme. As the initial guess, we use the $D_a = 0$ limit, where the dynamics reduce to Eq.\,(\ref{eq:dedimensionalisedeqequations}) and can be solved exactly. This solution is then substituted into Eq.\,(\ref{eq:Candchinoic}) and iterated until convergence. To proceed, we first discretize Eq.\,(\ref{eq:Candchinoic}) on a grid consistent with that of the initial guess. This grid is not uniform: its spacing doubles at fixed intervals, giving it a logarithmic-like structure. On this grid, the derivatives can be discretized as
\begin{equation}
    \dot{C}_i = \frac{1}{\delta_i}\left[ \frac{\alpha_i +1}{\alpha_i} C_i - \frac{\alpha_i}{\alpha_i - 1}C_{i-1} + \frac{1}{\alpha_i(\alpha_i-1)}C_{i-2}\right] + \mathcal{O}(\delta_i^3)
\end{equation}
where $\delta_i = \tau_i - \tau_{i-1}$ and $\alpha_i = \frac{\tau_i - \tau_{i - 2}}{\delta_i}$. For a linearly spaced grid $\alpha_i = 2$, and this formula reduces to Eq.\,(\ref{eq:derivativediscretisation}). We can discretize the integrals appearing in Eq.\,(\ref{eq:Candchinoic})
\begin{subequations}
    \begin{align}
        I^{(1)}(\tau) &\equiv \int_0^\tau d\tau' \sigma(\tau-\tau') \frac{\partial\chi(\tau-\tau')}{\partial \tau'}C(\tau) \\
        I^{(2)}(\tau) &\equiv \int_0^\tau d\tau' \sigma(\tau-\tau') \frac{\partial\chi(\tau-\tau')}{\partial \tau'}\chi(\tau) \\
        I^{(3)}(\tau) &= \int_0^{\infty} d\tau' m(\tau + \tau') \frac{\partial \chi(\tau')}{\partial \tau'} \\
        I^{(4)}(\tau) &= \int_{0}^{\infty} d\tau' \sigma(\tau + \tau') \frac{\partial \chi(\tau+\tau')}{\partial \tau'} C(\tau')
     \end{align}
\end{subequations}
using the approximation given in Eq.\,(\ref{eq:integralapprox}) and a linear interpolation, sufficient at this level as the functions are slowly-varying in time. Altogether, the error made in our approximation will be of order $\mathcal{O}(\delta^2 \frac{d^2 C}{d \tau^2})$. As before, we have used Simpson integration (here, with a variable time-step) to evaluate the bond integrals. This can be easily obtained by generalising the original method. Defining $\delta_j = \tau_j - \tau_{j-1}$ and $\alpha_j = (\tau_j - \tau_{j-2})/\delta_j$, we have 
\begin{equation}
    df_j = \frac{1}{6}\left[ - \frac{1}{\alpha_j(\alpha_j - 1)}f_{i-2}  + \frac{3\alpha_j - 2}{\alpha_j - 1}f_{i-1} + \frac{3\alpha_j - 1}{\alpha_j}f_j \right]
\end{equation}
Once again, setting $\alpha_j = 2$ for a linearly spaced grid, we recover Eq.\,(\ref{eq:simpsonrule}).

Putting all of these together, the functions $C$ and $\chi$ satisfy the following equations at $\tau = \tau_i$
\begin{subequations}
    \begin{gather}
        C_i = \frac{\dfrac{\alpha_i}{\delta_i(\alpha_i - 1)}C_{i-1} - \dfrac{1}{\delta_i \alpha_i(\alpha_i-1)}C_{i-2} + I^{(1)}_i - I^{(3)}_i - I^{(4)}_i}{\dfrac{\alpha_i + 1}{\delta_i \alpha_i} + 1 - pI^{(3)}_0} \\
        \chi_i = \frac{\dfrac{\alpha_i}{\delta_i(\alpha_i - 1)}\chi_{i-1} - \dfrac{1}{\delta_i \alpha_i(\alpha_i-1)}\chi_{i-2} + I^{(2)}_i -1}{\dfrac{\alpha_i + 1}{\delta_i \alpha_i} + 1 - pI^{(3)}_0}
    \end{gather}
\end{subequations}
where we have used a transparent indexing notation defining for a physical quantity $Q$, $Q_i \equiv Q(\tau_i)$. We iterate this equation until we get convergence to the desired solution. Specifically, we iterate until 
\begin{equation}
    \epsilon = \max\bigg( \max  \left( \left|C_i^{(n)} - C_i^{(n-1)} \right|\right),  \max  \left( \left|\chi_i^{(n)} - \chi_i^{(n-1)} \right|\right) \bigg)
\end{equation}
drops below $10^{-5}$, where the superscript $(n)$ denotes the number of iteration. Other metrics can also be defined such as 
\begin{equation}
    \epsilon = \frac{1}{2} \left(  \sqrt{\frac{\sum_{i = 1}^{N} \left[C^{(n)}_i - C^{(n-1)}_i\right]^2}{\sum_{i = 1}^{N} \left[C^{(n)}_i \right]^2 }} + \sqrt{ \frac{\sum_{i = 1}^{N} \left[\chi^{(n)}_i - \chi^{(n-1)}_i\right]^2}{\sum_{i = 1}^{N} \left[\chi^{(n)}_i \right]^2 } }\right)
\end{equation}

Since the correlation function and integrated response are bounded by $\pm 1$, we use an absolute rather than a relative difference as our convergence criterion. Relative differences become unreliable when comparing values close to zero.

To explore the parameter space, we proceed as follows: for fixed $D_a$ and $t_a$, we start at high temperature ($T \sim 0.7$) using the equilibrium solution as the initial guess, and then gradually decrease $T$ once convergence is reached. This slow cooling significantly improves convergence. In practice, we lower the temperature in steps of $0.005$, which allows us to probe the physics near the glass transition and resolve the two-step decay profiles.


%